\documentclass[pdflatex,sn-mathphys-num]{sn-jnl}

\usepackage{graphicx}%
\usepackage{multirow}%
\usepackage{amsmath,amssymb,amsfonts}%
\usepackage{amsthm}%
\usepackage{mathrsfs}%
\usepackage[title]{appendix}%
\usepackage{xcolor}%
\usepackage{textcomp}%
\usepackage{manyfoot}%
\usepackage{booktabs}%
\usepackage{algorithm}%
\usepackage{algorithmicx}%
\usepackage{algpseudocode}%
\usepackage{listings}%
\usepackage{siunitx}%
\usepackage{subcaption}

\theoremstyle{thmstyleone}%
\theoremstyle{thmstyletwo}%

\theoremstyle{thmstylethree}%

\begin{document}

\title[Particle tracking with physics-informed deep learning methods]{Particle tracking with physics-informed deep learning methods}

\author*[1,2]{\fnm{Matthias} \sur{Remta}}\email{matthias.remta@cern.ch}
\author[3]{\fnm{Anja} \sur{Beck}}\email{anbeck@mit.edu}
\author[4]{\fnm{Shanthalakshmi} \sur{Kilambi}}\email{s.kilambi@tudelft.nl}
\author[5]{\fnm{Thomas} \sur{Zhang}}\email{thzh2950@student.su.se}
\author[1]{\fnm{Francesco} \sur{Velotti}}\email{francesco.maria.velotti@cern.ch}


\affil[1]{\orgdiv{Systems}, \orgname{CERN}, \postcode{1211}, \city{Geneva}, \country{Switzerland}}

\affil[2]{\orgdiv{Faculty of Mathematics}, \orgname{University of Vienna}, \orgaddress{\street{Oskar-Morgenstern-Platz 1}, \postcode{1090}, \city{Vienna}, \country{Austria}}}

\affil[3]{\orgdiv{Department of Physics}, \orgname{MIT}, \postcode{02139}, \city{Cambridge}, \country{MA, USA}}

\affil[4]{\orgdiv{Faculty of Aerospace Engineering}, \orgname{Delft University of Technology}, \orgaddress{\street{Kluyverweg 1}, \city{Delft}, \postcode{2629HS}, \country{The Netherlands}}}

\affil[5]{\orgdiv{Department of Physics}, \orgname{Stockholm University}, \orgaddress{\street{Roslagstullsbacken 21}, \postcode{10691}, \city{Stockholm}, \country{Sweden}}}


\abstract{Simulating the motion of charged particles in electromagnetic fields is essential for designing and optimising particle accelerators. Conventional tools rely on symplectic integration schemes, which provide high accuracy but are computationally expensive. As a consequence, optimisation in moderate to high-dimensional parameter spaces as well as simulations of tens of thousands to millions of particles can be computationally prohibitive. 
This contribution explores the possibilities of employing modern machine-learning based tools, in particular SympNet and DeepONet, to enable fast particle simulations.
A major novelty is the modification of the conventional SympNet architecture to enable learning of parametric Hamiltonian dynamics.
The models are trained and tested on a toy setup of a circular accelerator comprising two different types of quadrupole magnets with varying field strengths.
All models achieved faster inference than the symplectic integrator, at the expanse of significantly reduced accuracy. The SympNet implementation achieved the lowest mean squared error. 
Additionally, a DeepONet was employed to predict the evolution of particle densities, derived from the single-particle simulations.
}




\maketitle

\section{Introduction}\label{sec:introduction}
\begin{figure}
    \begin{subfigure}{0.5\textwidth}
    \centering
        \includegraphics[width=\textwidth]{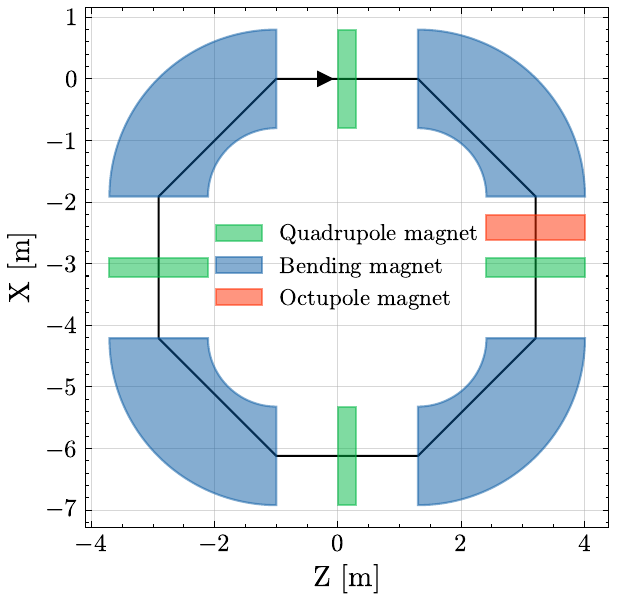}
    \end{subfigure}
    \hfill
    \begin{subfigure}{0.5\textwidth}
    \centering
        \includegraphics[width=\textwidth]{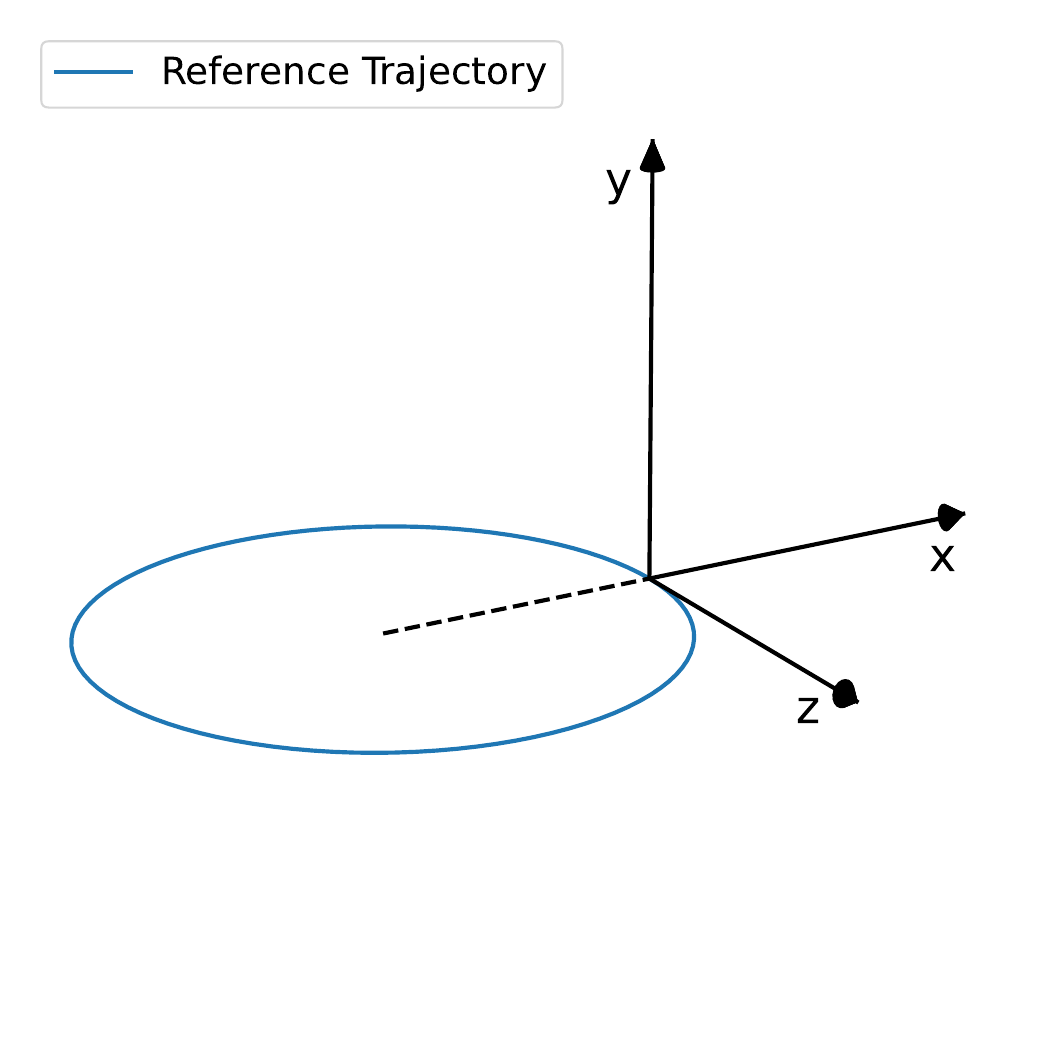}
    \end{subfigure}
    \caption{Left: top-down view of the simulated circular accelerator. Right: Frenet-Serret coordinate-system for a circular reference trajectory.}
    \label{fig:accelerator_schematic}
\end{figure}

The core components of a particle accelerator are the electromagnetic elements that guide and focus the beam as illustrated for example in the left panel of Fig.~\ref{fig:accelerator_schematic}. Simulating the motion of charged particles in the fields of these elements, referred to as particle tracking, is essential in the design, operation, and optimisation of modern particle accelerators. Most commonly, the motion of each individual particle is described as a Hamiltonian system \(\mathcal{H}(x, y, z, p_x, p_y, p_z, t)\) depending on the position of the particle, \(x, y, z\), its momenta, \( p_x, p_y, p_z\), and the time, \(t\). The time dependence arises from the different electromagnetic fields of the accelerator elements. Furthermore, the field of individual elements may vary over time to satisfy operational or experimental requirements, adding another layer of time dependence.
Instead of using a global coordinate system, it is usually convenient to express the position and momentum of each particle with respect to a predefined reference trajectory as illustrated for a perfectly circular accelerator in the right panel of Fig.~\ref{fig:accelerator_schematic}.

To date, many different particle-tracking tools have been developed, see e.g. Refs.~\cite{Grote:2003ct, Iadarola:2023fuk, kaiser2024cheetah}.
While the implementation details vary widely, most of them rely on the construction of symplectic -- or at least volume-preserving -- transformations to advance initial phase-space coordinates in time.
Modern particle accelerators however require simulations of up to millions of particles at once over timespans covering thousands of turns.
As a consequence, design studies and optimisation processes may be limited by the amount of available computing power, especially for moderate to high-dimensional parameter spaces. This contribution explores deep-learning methods as a faster, low-fidelity alternative.
Specifically, SympNet \cite{jin2020sympnets}, an autoregressive Fourier Neural Operator (FNO)~\cite{li2021, pathak2022} and DeepONet~\cite{Lu2021} are investigated.
These architectures are applied to trajectories of individual particles in a circulator accelerator with non-linear fields. In addition, DeepONet is applied to the evolution of the phase-space density, derived from the ensemble of individual particles. 

In recent years, there has been a surge in interest in deep-learning methods in particle physics \cite{Aad_2023, Aad_2025, jahin2025physicsinformedgraphneuralnetworks}. One prominent example is track reconstruction in detectors \cite{HEPTrkX, ju2020graphneuralnetworksparticle, THOMADAKIS2022108360, correia2024}.
Note that track reconstruction is sometimes also referred to as particle tracking but the problem setup is different to the one tackled in this paper.
At the same time, deep-learning methods remain largely underexplored in accelerator physics.
In the context of linear accelerators, FNOs have been employed to approximate statistical beam parameters obtained from simulations and a DeepONet has been shown to successfully predict for example the neutron flux, the production of secondary particles, and temperature changes at the experimental site~\cite{particles8010021}.
Moreover, a DeepONet has also been applied to neutron transport in nuclear-energy systems~\cite{Kobayashi2024}.
Ref.~\cite{PhysRevAccelBeams.23.074601} presents a neural network architecture based on the Taylor expansion of the transfer map of a particle accelerator and shows how symplectic conditions can regularise the training loss.
Finally, symplectic deep-learning methods based on H{\'e}non maps -- originally developed for plasma simulations~\cite{Burby_2021} -- have been applied to particle tracking in a linear accelerator~\cite{Huang_2024}.
This publication reaches beyond existing literature in two ways. First, neural operators are applied to circular accelerators with time-varying settings. Circular machines typically require tracking over many turns, resulting in significantly longer integration times compared to linear accelerators. Second, the SympNet architecture is extended to parametric Hamiltonian systems, enabling its application to multiple accelerator configurations with different time-dependency. 

The remainder of the paper is structured as follows: the setup -- including data and methods -- is described in Sec.~\ref{sec:methods}, the results are presented in Sec.~\ref{sec:results}, followed by a discussion and the conclusion in Sec.~\ref{sec:discussion_conclusion}.

\section{Setup}\label{sec:methods}
This section starts with a description of the toy accelerator setup used for the following studies. This includes the process for generating training and testing data representing particle motion in the accelerator.
Afterwards, the basic concepts defining a SympNet, a Fourier Neural Operator, and a DeepONet are discussed.

All data for this study is obtained with the simulation library \texttt{Xsuite} using a drift-kick scheme for symplectic integration~\cite{Iadarola:2023fuk}.
The simulated accelerator setup is displayed in Fig.~\ref{fig:accelerator_schematic}.
The phase-space is reduced to one spacial dimension to speed-up simulations. The positions and momenta for \num{e4} particles are initialised using a Gaussian distribution in phase-space. Afterwards, the particles are propagated through the accelerator for \num{e3} turns while phase-space coordinates are recorded after every turn. This procedure corresponds to following the particle motion on a Poincaré section taken at the periodicity of the accelerator. The simulations were performed for five different configurations of the accelerator -- labelled from A to E -- each starting with identical initial settings in the magnetic elements, followed by a linear ramp-up of the quadrupole field strengths which reach different final values. The field strength of a quadrupole can be described by the first-order normalised multipole strength \(k_1\).The simulated accelerator houses one family of focusing and one family of defocusing quadrupoles, each being powered in series. Therefore, the accelerator settings at any given time are characterised by two parameters,  \(k_1^f\) and \(k_1^d\), and a configuration by the time-evolution of these parameters.
Configurations A, B, C are used as training data and D, E are kept for testing.
Furthermore, phase-space densities are estimated from the ensemble of individual particles with Gaussian kernel density estimation (KDE) at 1024 collocation points per density. Collocation points are drawn with scrambled Sobol sampling \cite{OWEN1998466}.

One architecture considered in this paper are SympNets, a class of hard-constrained physics-informed neural networks. Because each layer is constructed as a symplectic transformation, the model, as a composition of symplectic transformations, is itself symplectic. In this work, linear layers \(\mathcal{L}\) of alternating upper and lower triangular matrices and activation layers \(A_{up}\) and \(A_{low}\) are used, following the LA-SympNet design presented in Ref.~\cite{jin2020sympnets}:

\begin{align}
\mathcal{L}_n\begin{bmatrix}
y \\
p_y 
\end{bmatrix} 
&=
\begin{bmatrix}
I & 0 / S_n \\
0 / S_n & I
\end{bmatrix}
\cdots
\begin{bmatrix}
I & 0 / S_2 \\
0 / S_2 & I
\end{bmatrix}
\begin{bmatrix}
I & 0 / S_1 \\
0 / S_1 & I
\end{bmatrix}
\begin{bmatrix}
y \\
p_y 
\end{bmatrix}
+ b \ , \\
A_{low} \begin{bmatrix}
y \\
p_y 
\end{bmatrix} 
&= 
\begin{bmatrix}
y \\
 \text{diag}(a)\sigma(y) + p_y
\end{bmatrix} \ , \\
A_{up} \begin{bmatrix}
y \\
p_y 
\end{bmatrix} 
&= 
\begin{bmatrix}
\text{diag}(a)\sigma(p_y) + y \\
p_y
\end{bmatrix} \ .
\end{align}
Symplecticity is ensured by choosing symmetric matrices in the linear layers via \(S_i = M_iM_i^T\).
This architecture cannot learn parametric Hamiltonian systems, as its input is restricted to phase space coordinates. The work presented later in this paper extends the SympNet architecture by adding additional networks to each layer, connecting the parameters of the Hamiltonian system to the weights of the symplectic transformation. Explicitly, if \(k \in \mathbb{R}^p\) denotes the \(p\)-dimensional parameter of the Hamiltonian, a linear layer \(\mathcal{L}_n\) is endowed with \(n\) neural networks \(\phi_i: k \mapsto M_i\). For each activation layer, a single neural network \(\phi: k \mapsto a\) suffices. The choice of architecture for those networks is unrestricted - in this paper multi-layer perceptrons are used.

Another architecture employed in this paper is the Fourier Neural Operator which maps input to output functions via linear transformations and convolutions with trainable kernels~\cite{li2021}. First, all input functions have to be discretised to an equidistant grid. Then additional input channels are added through a pointwise lifting operation. Finally, the data is passed through \(T\) Fourier layers, performing the transformation
\begin{equation}
    v_{t+1} = \sigma\bigg(Wv_t(x) + \mathcal{F}^{-1}\left(W_{\phi}(\mathcal{F}v_t)\right)(x)\bigg), \hspace{0.2cm} t=0, \dots, T \ .
\end{equation}
Here, \(\mathcal{F}\) denotes the discrete Fourier transformation, \(\sigma\) is an activation function, \(W\) a pointwise-applied linear transformation and \(W_{\phi}\) the discrete Fourier transform of a periodic function. Hence, the convolution is parametrised directly in Fourier space with trainable coefficients \(W_{\phi}\). 
In this paper, FNO is applied as an autoregressive model, following Ref.~\cite{pathak2022}.

The third deep-learning tool investigated in this paper is the DeepONet, which is able to learn operators \(\mathcal{G}: u \mapsto G(u)\). Inspired by the universal approximation theorem for operators \cite{chen1995}, a DeepONet \cite{Lu2021} is comprised of two neural networks: the branch net and the trunk net. The branch net takes a discretisation \(\left[u(x_1), \dots, u(x_m)\right]\) of the function \(u\) at fixed sensors as input and outputs a vector \(b \in \mathbb{R}^d\). The input to the trunk net is the location \(z\) where the transformed function \(G(u)\) should be evaluated. The output of the trunk net is a vector \(v \in \mathbb{R}^d\). The dimension \(d\) of the internal representations \(b\) and \(v\) can be freely chosen. Finally, the value of the transformed function \(G(u)\) at the location \(z\) is approximated by the dot product:
\begin{equation}
    G(u)(z) \approx \sum\limits_{i=1}^{d} b_iv_i .
\end{equation}
Note that in this work, the branch and trunk nets consist of ResNet~\cite{7780459} architectures with linear layers instead of convolution layers.

\section{Results}\label{sec:results}
This section discusses the possibilities and constraints of the previously introduced tools in the context of particle tracking.
This includes the prediction of individual particle trajectories in the first part of this section and the prediction of the phase-space density defined by the ensemble of the particles in the second part of this sections.
This section concludes with a discussion of the speed-up achieved by each method.
\subsection{Particle trajectories}
All three models, SympNet, FNO and DeepONet, are trained to predict trajectories of individual particles using the same training data with specific preprocessing steps depending on the architecture. The training data set consists of \num{3e5} samples. In order to speed up the training, the phase-space is reduced to one spacial dimension, i.e. \(\vec{\Phi}(t)=(y(t), p_y(t))\).

\begin{figure}
    \centering  
    \includegraphics[width=1.0\textwidth]{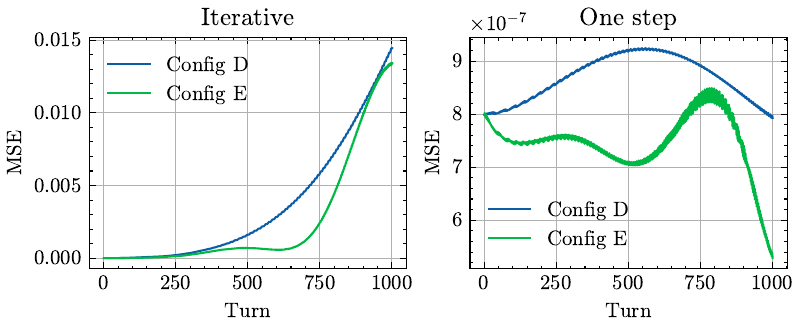}
    \caption{Left: MSE achieved by the SympNet when applied iteratively to integrate \num{e3} turns. Right: one-step errors with the true phase space coordinates as inputs. MSE errors in both panels are reported for normalised coordinates.}
    \label{fig:sympnet_loss}
\end{figure}

\begin{figure}
    \centering  
    \includegraphics[width=1.0\textwidth]{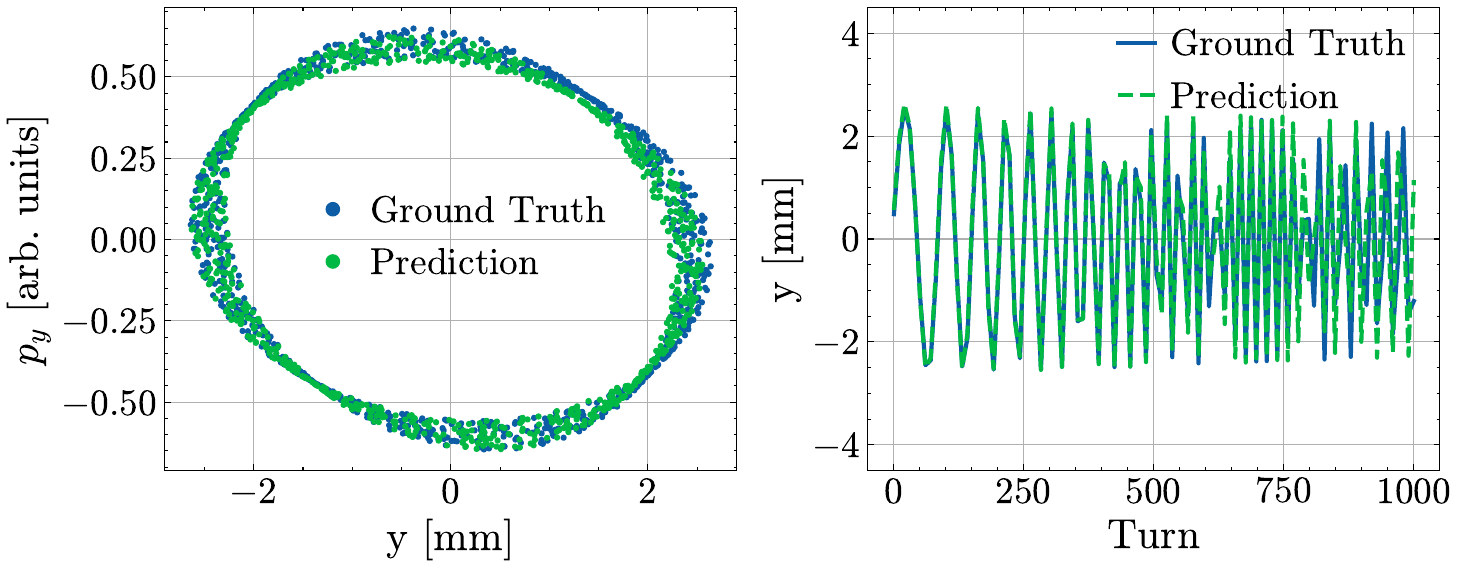}
    \caption{The evolution of phase-space coordinates over 1000 turns is predicted by iteratively applying SympNet, starting from one exemplary initial condition. Left: Predicted phase-space coordinates versus ground truth on the Poincaré section. Right: Prediction and ground truth of the position for the same initial condition. Every tenth turn is shown for visual clarity.}
    \label{fig:sympnet_example_trajectory}
\end{figure}

The SympNet model is presented with phase-space coordinates \(\vec{\Phi}(t)\) and accelerator settings \(k_1^f(t), \ k_1^d(t)\) as inputs, with the coordinates $\vec{\Phi}(t+1)$ after one turn as targets. At inference, the model is iteratively applied to the initial phase-space coordinates, hence serving as an integrator. The evolution of errors over many integration steps is a key figure of merit, reported alongside the error of single integration steps in Fig.~\ref{fig:sympnet_loss}. Here, the error between model predictions and ground-truth phase-space coordinates is measured with the mean squared error (MSE), defined as the coordinate-wise squared difference averaged over all coordinates and particles. While the MSE for one time-step is very small with an order of \num{e-7}, these errors accumulate over multiple integration steps (see left panel of Fig. \ref{fig:sympnet_loss}), leading to significant deviations from the ground truth phase-space coordinates after many turns. The left panel of Fig.~\ref{fig:sympnet_example_trajectory} shows turn-by-turn phase-space coordinates for one initial condition of the ground truth data and the iterative model predictions for the same initial condition. These points trace out a similar area in the Poincaré section, showing that the energy is well approximated. Nonetheless, the right panel of Fig.~\ref{fig:sympnet_example_trajectory} reveals significant deviations in the spacial coordinate, caused by the phase error. As the energy is preserved by the symplectic architecture, the MSE is overall bounded.  

\begin{figure}
    \centering  
    \includegraphics[width=1.0\textwidth]{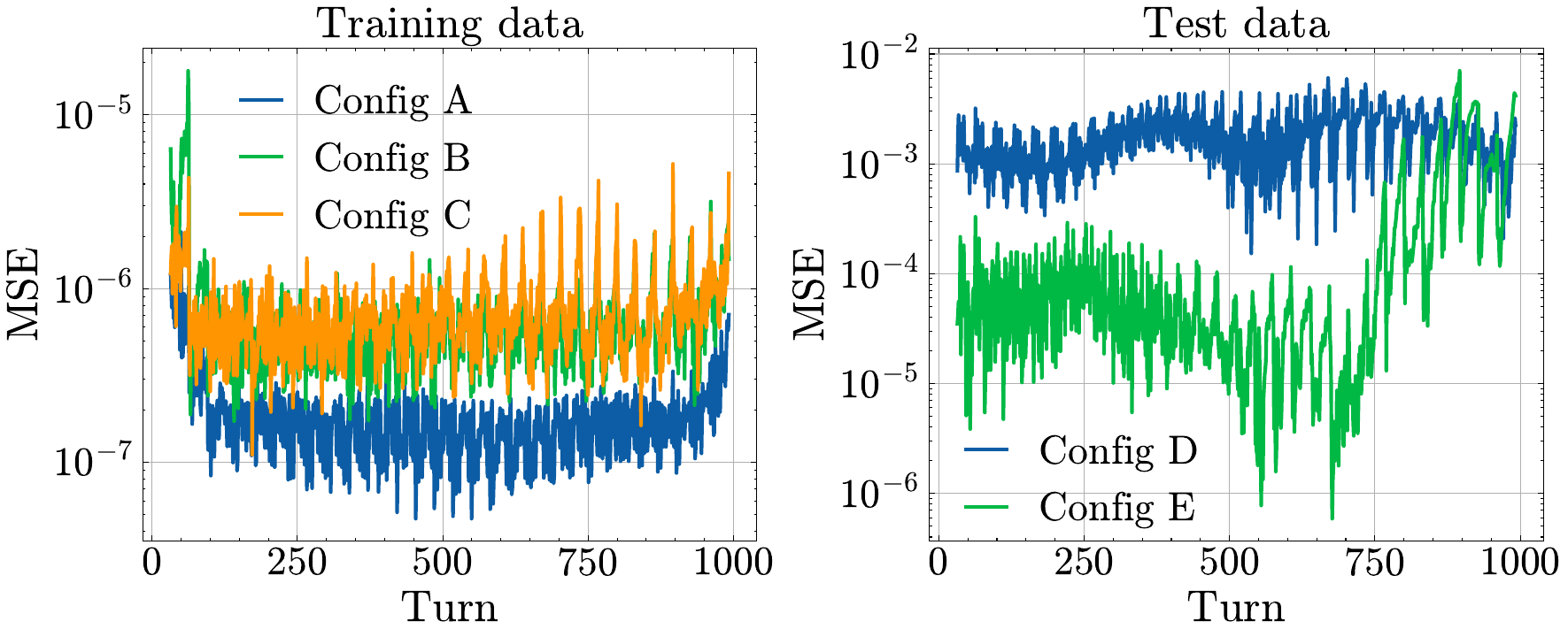}
    \caption{Left: MSE for FNO predicting a single time-step on the training data. Right: MSE for FNO predicting a single time-step on the test data. MSE errors in both panels are reported for normalised coordinates.}
    \label{fig:fno_one_step_loss}
\end{figure}

\begin{figure}
    \centering  
    \includegraphics[width=1.0\textwidth]{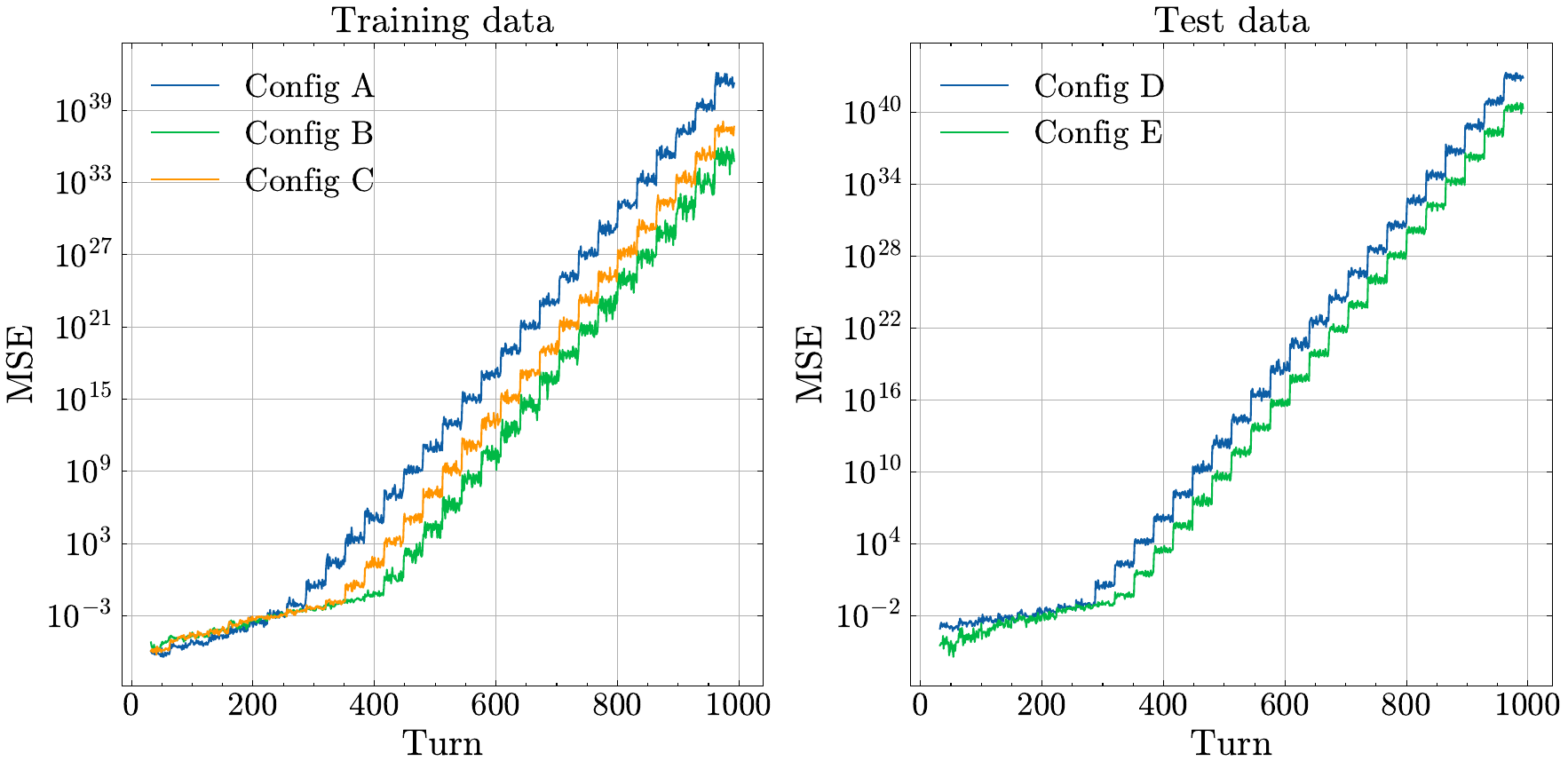}
    \caption{Left: MSE for FNO predicting the training data iteratively. Right: MSE for FNO predicting the test data iteratively. MSE errors in both panels are reported for normalised coordinates. The step-like shape of the curves originates from the model predicting 32 timesteps per iteration.}
    \label{fig:fno_iterative_loss}
\end{figure}

The FNO is applied as an integrator, taking the phase-space coordinates \(\vec{\Phi}\) of the last \(n\) turns and a discrete representation of the time-evolution of the settings \(k_1^f, \ k_1^d\) as input and predicting the subsequent \(n\) turns in a single integration step. For training and inference \(n\) is set to 32. The model is exclusively trained to perform single integration steps on ground truth coordinates as input. In order to obtain predictions for more steps, the model is deployed auto-regressively at the inference stage. The resulting errors for a single prediction step are shown in Fig.~\ref{fig:fno_one_step_loss}. The MSE for the training data is in the order of \numrange{e-7}{e-5} (left panel) and in the range of \numrange{e-5}{e-3} for the test data (right panel).
Figure~\ref{fig:fno_iterative_loss} reveals an exponential growth of the MSE. Unlike SympNet, the FNO is not equipped with an architecture that preserves invariants, hence the errors grow without bounds.

The DeepONet is employed as a surrogate for the operator \mbox{\(\mathcal{G}: (y(0), p_y(0); k_1^f, k_1^d) \mapsto (y, p_y)\)}, which maps initial conditions to the corresponding phase-space trajectories. The branch net is provided with initial phase-space coordinates at \(t=0\) and the time-evolution of the quadrupole strengths \(k_1^f, \ k_1^d\), captured on a grid of equidistant sensor. To facilitate learning the quasi-periodical motion, the time is provided with an additional Fourier expansion to the trunk net. Because the output of the function \((y, p_y)\) is two dimensional, we choose an even number \(d=64\) for the dimension of the internal representations and set:
\begin{equation}
    G\left(y(0), p_y(0);  k_1^f, k_1^d\right)(t) \approx \left(\sum\limits_{i=1}^{d/2} b_iv_i, \sum\limits_{i=d/2 + 1}^{d}b_iv_i\right) \ .
\end{equation}

\begin{figure}
    \centering  
    \includegraphics[width=1.0\textwidth]{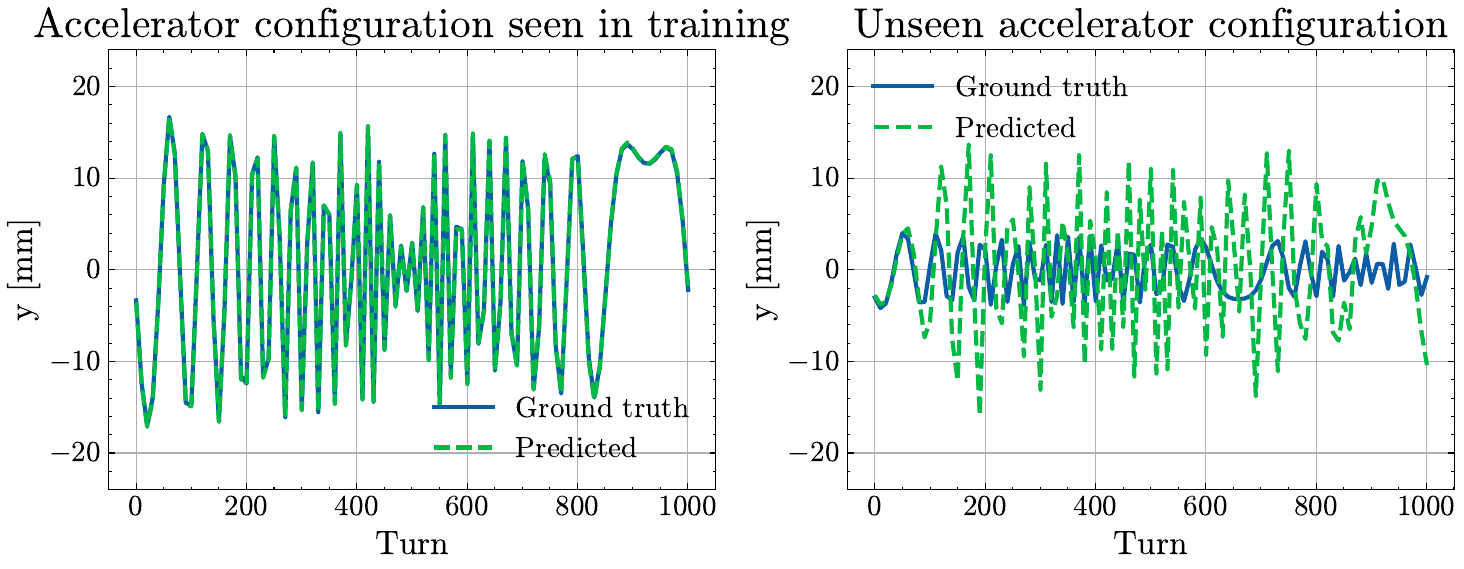}
    \caption{Performance of the DeepONet on training and test data. Left: representative example trajectory for an accelerator configuration included in the training set. The model inputs include unseen combinations of initial conditions and time. Right: representative example trajectory for an accelerator configuration not included in the training set. The model fails to accurately predict trajectories in unknown configurations.}
    \label{fig:deeponet_example_trajectories}
\end{figure}

In our experiments, the DeepONet training does not converge when using all \num{e3} turns. A possible cause could be the spectral bias of neural networks~\cite{rahaman2019spectralbiasneuralnetworks}. As the particle motion exhibits high-frequency components with respect to the normalised time, convergences rates can become prohibitively slow. One solution is to only include every tenth turn in the training data set, effectively aliasing the higher frequencies. In this case, the DeepONet performs well as a surrogate for accelerator configurations that were included in the training set. It accurately predicts phase-space coordinates for combinations of initial conditions and time that it did not see during training as illustrated by the perfect agreement between truth and prediction in the left panel of Fig.~\ref{fig:deeponet_example_trajectories}. The maximum Euclidean distance is \num{1e-1} in normalised coordinates and the median is \num{3e-3}. However, the performance degrades significantly when attempting to predict unseen accelerator configurations. The right panel of Fig.~\ref{fig:deeponet_example_trajectories} shows a representative mismodelled trajectory. In this case the maximum Euclidean distance is \num{3.0} and the median Euclidean distance is \num{6e-1}. In conclusion, the model would not be useable as a surrogate.

\subsection{Phase-space densities}\label{sec:densities}
\begin{figure}
    \centering
    \begin{subfigure}{\linewidth}
        \centering
        \includegraphics[width=1.0\linewidth]{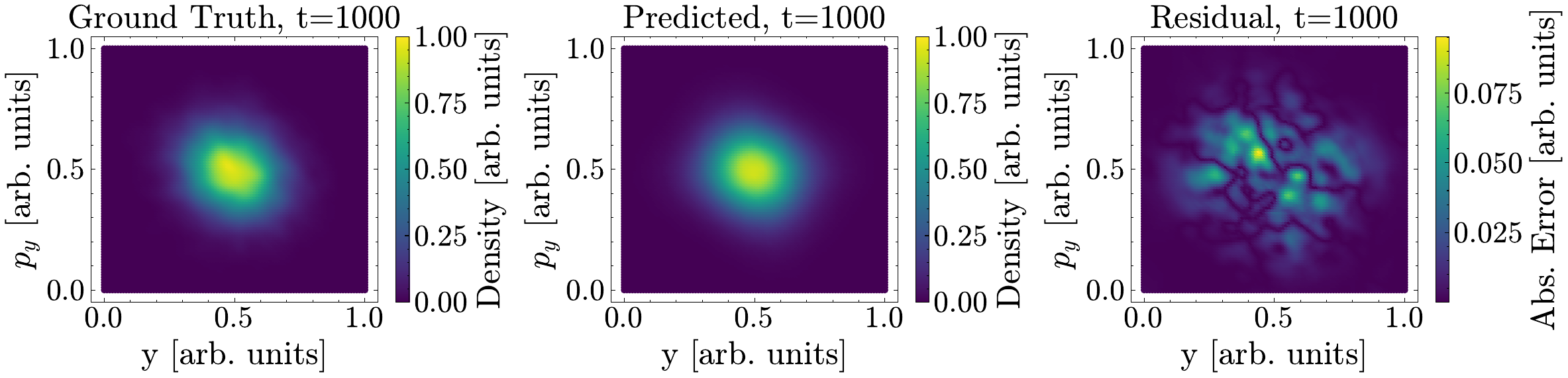}
    \end{subfigure}

    \vspace{1em} 

    \begin{subfigure}{\linewidth}
        \centering
        \includegraphics[width=1.0\linewidth]{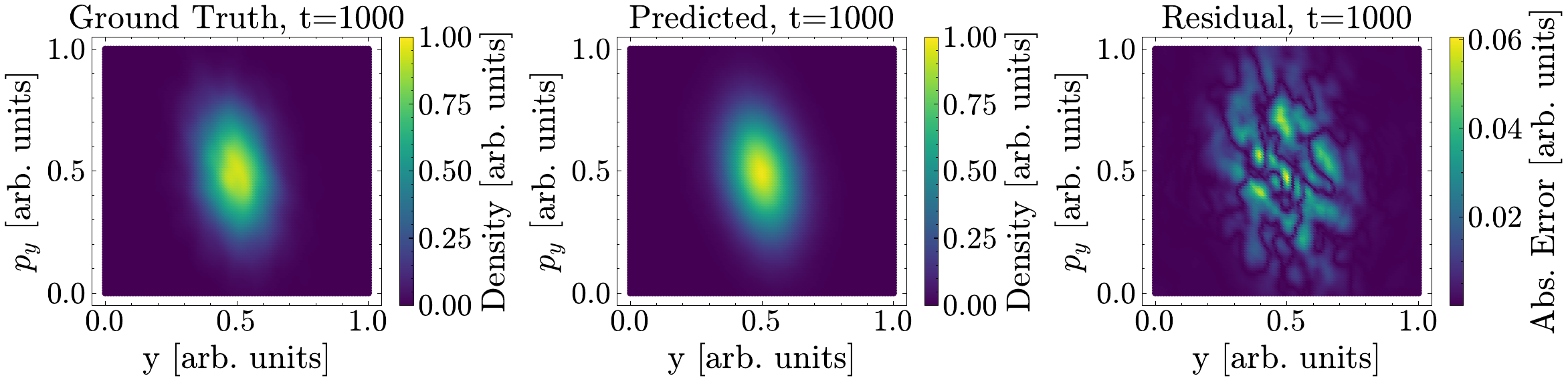}
    \end{subfigure}

    \caption{Left: Ground truth of the phase space densities after 1000 turns for the two accelerator configurations D, E (top/bottom) in the test set. Center: Model prediction. Right: Absolute error between ground truth and model prediction. The densities are evaluated on a 96x96 equidistant grid.}
    \label{fig:deeponet_example_densities}
\end{figure}

\begin{figure}
    \centering  
    \includegraphics[width=1.0\textwidth]{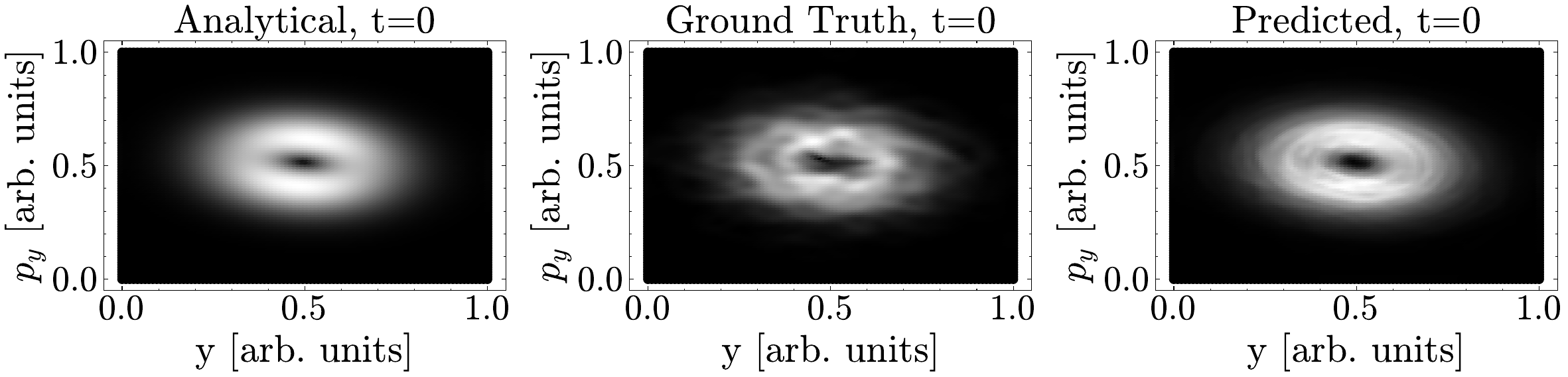}
    \caption{Left: Gradient of the analytical distribution used to generate the ground truth. Center: The ground-truth density, estimated with Gaussian KDE from an ensemble of particles. Right: The predicted density at \(t=0\). Gradients were estimated by applying Sobel filters \cite{sobel1973} to an equidistant 96x96 grid.}
    \label{fig:deeponet_density_gradients}
\end{figure}

In this section we turn our attention towards the time-evolution of phase space densities under the Hamiltonian system. Given an initial density \(\rho_0\), we surrogate the Frobenius-Perron operator \(\mathcal{P}_t\): \(\rho_0 \mapsto \rho_t\)~\cite{Lasota1994}.
Compared to individual particles, the phase-space density evolves much more slowly. The experiments discussed in the following start with particle distributions in a Gaussian equilibrium, as commonly seen in real-world accelerators, and slowly evolve by changing the quadrupole strengths. This allows to circumvent the spectral bias seen on the per-particle-level by training surrogate models on the density level. For this study, we choose a DeepONet where the branch net receives the information on the accelerator elements, and the trunk net receives the coordinates in the two-dimensional phase-space defined by \((y, p_y)\) and the time at which to evaluate the density. The predicted phase-space density after 1000 turns for the two accelerator configurations D, E in the test set are shown in the top and bottom rows of Fig.~\ref{fig:deeponet_example_densities}, alongside their respective ground truth and the absolute error. The ground truth densities exhibit single peaks in the two-dimensional phase-space which are captured very well by the model with non-systematic deviations. 
Because the initial conditions are drawn from a Gaussian distribution, we can compare the model predictions not only to the ground truth represented by an ensemble of individual particles but also to the analytical density for \(t=0\). Figure~\ref{fig:deeponet_density_gradients} reveals that the Gaussian KDE from a finite number of particles induced non-physical distortions in the density. The operator learned a smoother approximation, which resembles the analytical density more closely. In this case, the effect is beneficial, as it essentially removes noise induced by numerical effects. However, in other cases micro-structures in the densities may originate from the dynamics of the system. Hence, one has to carefully monitor the behaviour of the model.

\subsection{Computation time}
Table~\ref{tab:computation_time} compares the computational cost for the conventional (Xsuite) and deep-learning based simulation tools presented previously.
The states times are the average values obtained from several independent test runs.
All models achieve faster inference times than the numerical integrator, with speed-ups ranging from a factor of four for the DeepONet predicting densities to a factor of 20 for the DeepONet predicting particle trajectories. If only a specific point \(t\) in time is of interest, e.g. the end of the trajectory after \num{1e3} turns, Xsuite,  SympNet and the autoregressive FNO need to perform all iterative steps up to \(t\). The DeepONet models on the other hand are capable of predicting the phase-space coordinates or densities at \(t\) in a single pass. This methodological advantage yields a speed-up of three orders of magnitude.

\begin{table}[htbp]
\centering
\begin{tabular}{ccc}
\toprule
Method & Inference Time (entire test set)  & Inference Time (only \(t\) = \num{e3})\\
\midrule
Xsuite                  & \SI{1073.7}{\second}  & \SI{1073.7}{\second}       \\
SympNet                 & \SI{189.51}{\second}   & \SI{189.51}{\second}       \\
FNO                     & \SI{111.51}{\second}   & \SI{111.51}{\second}       \\
DeepOnet (trajectories) & \hphantom{1}\SI{50.09}{\second}    & \hphantom{11}\SI{0.19}{\second}      \\
DeepOnet (densities)    & \SI{280.12}{\second}   & \hphantom{11}\SI{0.13}{\second}      \\
\bottomrule
\end{tabular}
\caption{Comparison of inference times of the numerical solver (Xsuite) and the deep-learning models for the entire test set. All evaluations where performed on an Intel Core i7-1355U CPU @ 1.70 GHz.}
\label{tab:computation_time}
\end{table}

\section{Discussion and Conclusion}\label{sec:discussion_conclusion}
The inference time of numerical models for particle accelerators is a key bottle-neck in the optimisation of existing, and the development of novel accelerators.
Real-time surrogate models for CERN's injectors are even more challenging due to the very high repetition rate paired with integration of many particles over large timescales.
This paper investigated the potential of different deep-learning surrogate models in terms of accuracy and speed-up.
A novelty developed in this publication is extending the standard SympNet architecture to handle parametric Hamiltonian systems.
All three studied architectures -- SympNet, FNO and DeepONet -- demonstrated a significant reduction in evaluation time, see Tab.~\ref{tab:computation_time}, with DeepONet achieving the fastest inference time.
Given the toy model studied here is orders of magnitude simpler than accelerators at CERN~\cite{1211818}, the speed-up in the real-world scenario is likely even more striking.
When applied to individual particles, only SympNet achieved acceptable accuracy on the test set.
The DeepONet was also applied to particle densities with excellent accuracy and significant speed-up over the numerical integration method.

The iterative models -- SympNet and FNO -- were trained with teacher forcing \cite{10.1162/neco.1989.1.2.270}. During inference, however, the FNO’s errors grew exponentially with each pass through the model. A future study could address this mismatch between training and inference by exposing the model to its own predictions during training \cite{NIPS2015_e995f98d}.

Benchmarking on real-world applications in the field of accelerator physics is a logical next step, to further evaluate the robustness and generality of the methods.
This includes in particular predictions of the full 6D phase space.

\backmatter

%
%
%

%
%

\section*{Declarations}

\subsection*{Funding}
Not applicable.

\subsection*{Conflict of interest}
The authors declare that they have no known competing financial interests or personal relationships that could have appeared to influence the work reported in this paper.

\subsection*{Ethics approval and consent to participate}
Not applicable.

\subsection*{Consent for publication}
Not applicable.

\subsection*{Data availability}
The datasets supporting the conclusions of this article are available in the GitHub repository, \url{https://github.com/MatthiasRemta/pinn-project.git}.

\subsection*{Materials availability}
Not applicable.

\subsection*{Code availability}
The code to reproduce the results reported in this paper is available at \url{https://github.com/MatthiasRemta/pinn-project.git}.

\subsection*{Author contribution}
M.R. conceived the idea, contributed to the numerical simulations, all the deep-learning studies and prepared the manuscript. \\
\noindent
A.B. contributed to the DeepONet studies predicting the particle density and the review of the document. \\
\noindent
S.K. contributed to the SympNet studies and the review of the document.\\
\noindent
T.Z. contributed to the Fourier Neural Operator studies and the review of the document.\\
\noindent
F.V. supervised the work of M.R. and contributed to the review of the document.
%


\bibliography{bibliography_revised}

\end{document}